\documentclass[conference]{IEEEtran}
\IEEEoverridecommandlockouts
\usepackage{cite}

\usepackage{amsmath,amssymb,amsfonts}
\usepackage{graphicx}
\usepackage{textcomp}

\usepackage{verbatim}
\usepackage{algorithm}
\usepackage{algpseudocode}
\usepackage{mathtools}
\usepackage[T1]{fontenc}
\usepackage[utf8]{inputenc}
\usepackage{lipsum}
\usepackage[table]{xcolor}
\usepackage{tabularx}
\usepackage{caption}
\usepackage{soul}
\usepackage[export]{adjustbox}
\usepackage{setspace}
\usepackage{multirow}

\def\BibTeX{{\rm B\kern-.05em{\sc i\kern-.025em b}\kern-.08em
    T\kern-.1667em\lower.7ex\hbox{E}\kern-.125emX}}

\renewcommand{\baselinestretch}{0.99}
    
\begin{document}

\title{Quantifying Uncertainty in Machine Learning-Based Pervasive Systems: Application to Human Activity Recognition
}

\author{\IEEEauthorblockN{ Vladimir BALDITSYN}
\IEEEauthorblockA{LCIS, Grenoble INP - UGA,\\ Valence, France \\
vladimir.balditsyn@imag.fr}
\and

\IEEEauthorblockN{Philippe LALANDA}
\IEEEauthorblockA{Univ. Grenoble Alpes,\\ Grenoble, France \\
philippe.lalanda@imag.fr}
\and
\IEEEauthorblockN{German VEGA}
\IEEEauthorblockA{Univ. Grenoble Alpes,\\ Grenoble, France \\
german.vega@imag.fr}
\and
\IEEEauthorblockN{Stéphanie CHOLLET}
\IEEEauthorblockA{LCIS, Grenoble INP - UGA,
\\ Valence, France \\
stephanie.chollet@esisar.grenoble-inp.fr}

}
\maketitle

\begin{abstract}

The recent convergence of pervasive computing and machine learning has given rise to numerous services, impacting almost all areas of economic and social activity. However, the use of AI techniques precludes certain standard software development practices, which emphasize rigorous testing to ensure the elimination of all bugs and adherence to well-defined specifications. ML models are trained on numerous high-dimensional examples rather than being manually coded. Consequently, the boundaries of their operating range are uncertain, and they cannot guarantee absolute error-free performance. In this paper, we propose to quantify uncertainty in ML-based systems. To achieve this, we propose to adapt and jointly utilize a set of selected techniques to evaluate the relevance of model predictions at runtime. We apply and evaluate these proposals in the highly heterogeneous and evolving domain of Human Activity Recognition (HAR). The results presented demonstrate the relevance of the approach, and we discuss in detail the assistance provided to domain experts.

\end{abstract}


\section{Introduction}

The recent convergence of pervasive computing and machine learning (ML) has led to the creation of numerous highly valuable and sophisticated services. This development is transforming almost every sector of economic and social activity \cite{pervasiveTrend}. For instance, it is particularly evident in the fields of industry, with the advent of Industry 4.0, healthcare, transportation, or smart residential environments. This integration is, therefore, making a significant impact on both businesses and everyday life.

However, the reliance on ML-based services introduces several fundamental challenges. In particular, it implies a major conceptual shift from traditional development methodologies, which emphasize rigorous testing as a crucial step to ensure the elimination of all bugs and adherence to well-defined specifications. ML models, on the other hand, are trained on examples (usually numerous and of high dimension) rather than being manually coded. As a result, the boundaries of their operating range are uncertain, making it impossible to formally prove that specific constraints are consistently met. 

This raises concerns about reliability, safety, and accountability, as ML-based solutions may behave unpredictably under certain conditions. They are particularly sensitive to noisy or inaccurate data, and more generally to differences between  the data distribution at training time and at inference time. The different types of distributional shifts are generally categorized as follows:

\begin{itemize}
\item Sample selection bias that occurs when the training data is not a representative sample of the population, leading to skewed model predictions. 
\item Covariate shift that occurs when the input feature distribution changes between training and deployment while their relationship with the target remains constant.
\item Label shift that occurs when the distribution of the labels changes between  training and deployment, while the class-conditional distributions remain constant.
\item Domain shift that occurs when models are applied to a different domain than they were trained
on, involving changes in both input features and target distributions. 
\end{itemize}

This sensitivity can lead to significant performance degradation, as the model may not generalize well to new, unseen data. Furthermore, small deviations in the data can accumulate and result in substantial errors, making it challenging to maintain the desired quality of service. It is therefore important to provide a level of uncertainty with every prediction,  especially in practical fields like healthcare or industrial automation, where wrong decisions can have significant negative consequences. Understanding the level of uncertainty helps users assess the reliability of a prediction and manage risks effectively. When uncertainty is high, decision-makers can proceed with caution or seek additional information before making a critical decision. Conversely, when uncertainty is low, decisions can be made more confidently. For instance, in predictive maintenance within Industry 4.0 \cite{1580414}, if a model predicts a malfunction with low uncertainty, operators can confidently plan maintenance. However, high uncertainty prompts operators to conduct additional tests or gather further evidence before committing to costly operations.

To address this issue, the field of uncertainty quantification (UQ) has seen significant development in recent years. In deep learning, different methods have been proposed and applied in various domains in order to associate a level of uncertainty with every prediction \cite{UQReview}. Some methods, such as Bayesian models, generate full distributions over predictions, allowing for the assessment of both likely outcomes and their variability. Others exploit data and neural network properties without relying on probabilistic models. For instance, some approaches seek to characterize the distribution or density of training data, often within the latent space, to position the inference data and determine the correspondence between training and runtime points. Although these approaches show promise, their effectiveness is often constrained to specific scenarios. For instance, while Bayesian methods offer strong theoretical foundations, they are computationally demanding and difficult to scale. Likewise, non-Bayesian methods that rely on assumptions about the latent space structure often fail to detect distribution shifts that maintain density or overall distribution but alter the semantic or functional properties of the data. These limitations compromise their reliability in real-world settings with unpredictable conditions.


The goal of this paper is to propose a novel approach that combines complementary uncertainty quantification techniques to address a variety of scenarios. The primary function of this framework is to assign a confidence flag, either green or red, to model predictions, where a green flag denotes low uncertainty, and a red flag indicates high uncertainty. The effectiveness of the proposed approach is demonstrated in the domain of smartphone-based Human Activity Recognition (HAR). More specifically, the contributions of this paper can be summarized as follows:

\begin{itemize}
    \item The development of a framework integrating uncertainty quantification (UQ) techniques for deep learning models. We chose a flag-based approach over percentage-based quantification, as the latter often lacks practical reliability and interpretability.
    \item The selection, adaptation, and implementation of UQ techniques designed to handle the various types of distributional shifts discussed above. These techniques include computing the reconstruction loss with an autoencoder, measuring the distance between runtime data and training data in the latent space, and applying Monte Carlo Dropout to generate multiple stochastic predictions.
    \item The implementation of this approach using a Transformer-based deep learning model, built with an initial unsupervised phase leveraging the Masked Autoencoders (MAE) technique.
    \item The design of a testing strategy encompassing various types of distributional shifts and the analysis of results for both individual techniques and their combined effectiveness.
\end{itemize}

The paper is organized as follows. First, we provide background on Human Activity Recognition and Uncertainty Quantification  in deep learning. Next, we introduce our proposal for combining various UQ techniques. This is followed by a detailed description of our approach’s implementation, including the HAR models and UQ methods employed. Section IV presents and discusses our experimental results, and the paper concludes with a summary of our findings.
\newpage

\section{Background}

\subsection{Human Activity Recognition}

We explore Human Activity Recognition (HAR), a field in which an activity is defined as a sequence or combination of simple motion patterns that exhibit temporal correlations \cite{bobick1997movement}. In our case, HAR is  performed using wearable devices, such as smartphones, which are equipped with Inertial Measurement Unit (IMU) sensors, including accelerometers, gyroscopes, and magnetometers. The prevailing approach to addressing this challenge relies on  machine learning techniques, which have become the standard for processing and interpreting sensor data in HAR applications.


The HAR domain frequently encounters significant challenges in achieving robust generalization due to various types of data shifts between training conditions and real-world testing scenarios. One major factor contributing to these discrepancies is the inherent variability in human behavior—individuals perform the same activities differently based on a range of personal attributes, such as age, gender, and physical fitness. Additionally, variations in hardware and software introduce further inconsistencies. Devices differ in terms of sensor types, operating systems, and hardware generations, leading to heterogeneous data distributions. As technology advances, older datasets become progressively outdated, reducing their applicability to newer devices and environments. Moreover, activity recognition models must adapt to the emergence of novel activities influenced by evolving user behaviors and changing environmental contexts. 


\begin{figure}[h!]
\centering        \includegraphics[width=6cm]{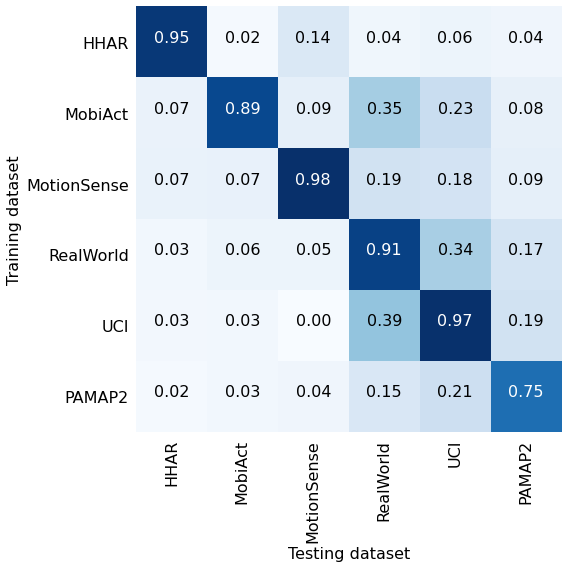}    \caption{Cross-dataset evaluation with popular datasets. A cell at coordinates $i,j$ shows the F1 score obtained by training the model with $70\%$ of dataset $i$ on $30\%$ of dataset $j$ \cite{presotto2023}.}
\label{fig:cross_dataset} \end{figure} 

Figure~\ref{fig:cross_dataset} presents the result of experiments showing the challenge of training a model on one dataset and executing it on a different one \cite{presotto2023}. It  demonstrates that the model performs well when trained and tested on the same dataset, but its performance significantly deteriorates in cross-dataset scenarios. In such cases, it is primordial to associate uncertainty with a prediction. There are many cases where predictions are wrong due to the intrinsic heterogeneity of the field. 

\subsection{Uncertainty Quantification}

Uncertainty is commonly categorized into two main types: aleatory and epistemic uncertainty \cite{hora1996,derkiureghian2009}. Aleatory uncertainty refers to the uncertainty that arises due to inherent randomness or noise in the data and is thus irreducible. Epistemic uncertainty refers to the uncertainty that arises from a lack of knowledge about the model or the data. It reflects the model’s ignorance about certain aspects of the problem due to insufficient training data, limited model capacity, or poor representation of the input space. Epistemic uncertainty can be mitigated with additional information or through the development of better algorithms.

A usual method for uncertainty estimation in neural networks uses softmax probabilities. The softmax function converts raw output scores (logits) into probabilities, interpreted as confidence levels. Hendrycks and Gimpel ~\cite{hendrycks2016baseline} proposed modeling uncertainties as a probability distribution over model parameters. Defined as $\sigma(\mathbf{z})_i = \frac{e^{z_i}}{\sum_j e^{z_j}}$, the softmax normalizes logits into a probability distribution, measuring uncertainty. However, this approach often overestimates confidence, even for incorrect predictions or out-of-distribution (OOD) data~\cite{hendrycks2017baseline}. This remains true even with improvements like the Inhibited Softmax, which adds an inhibition term to refine probabilities based on model confidence~\cite{mozejko2019inhibited}.

Bayesian methods \cite{hinton1993van, mackay1992} provide a probabilistic framework to interpret model parameters and quantify prediction uncertainty. By incorporating prior distributions and updating them with observed data, they yield posterior distributions that capture uncertainty. Approximations of these distributions include variational inference \cite{graves2011}, Monte Carlo dropout \cite{gal2016}, Markov Chain Monte Carlo \cite{christophe2023}, and model ensembles \cite{lakshminarayanan2017} \cite{neal2012, blundell2015}. Monte Carlo Dropout applies dropout at inference time to generate multiple predictions and estimate their variance. Model ensembles train multiple independent neural networks and aggregate their predictions to assess dispersion. However, these methods often rely on strong distributional assumptions, which can introduce inaccuracies when the true distribution deviates. Additionally, their computational complexity  may hinder their application in large  architectures.

Some studies rely on non-probabilistic approaches, using deterministic techniques to indirectly estimate confidence. For instance, temperature scaling of the softmax function calibrates output probabilities, mitigating excessive overconfidence \cite{Guo2017}. Gradient-based methods assess model sensitivity to input variations, identifying regions of uncertainty \cite{Goodfellow2015}. Autoencoders \cite{Liou2014} can also estimate uncertainty by analyzing variations in reconstruction loss across input data. While these techniques provide effective uncertainty estimates, they remain approximations, as they do not explicitly model uncertainty distributions. Moreover, lacking an inherent mechanism to distinguish unknown data, they are susceptible to errors with atypical or adversarial samples and do not naturally quantify their own uncertainty.

These methods yield good results in specific scenarios but lack generalizability. They are not universal techniques capable of handling different types of distribution shifts.

\newpage

\section{Proposal}

\subsection{Overview}

This study introduces a unified framework that seamlessly integrates complementary uncertainty quantification techniques to tackle a wide range of distributional shifts. Designed for the IoT domain, these techniques are fully implemented and readily available. Although initially applied to Human Activity Recognition, they can be easily adapted to other pervasive applications. 

At its core, the framework employs thresholding mechanisms to effectively distinguish in-distribution from out-of-distribution data. It combines three effective UQ strategies, each well suited to specific types of shift but limited when used in isolation:

\begin{itemize}
    \item Reconstruction loss analysis via an autoencoder to detect anomalies.
    \item Latent space distance measurement to track deviations between runtime and training distributions.
    \item Variance estimation across multiple stochastic predictions using Monte Carlo Dropout.
\end{itemize}


\begin{figure}[h!]
\centering        
\includegraphics[width=9cm]{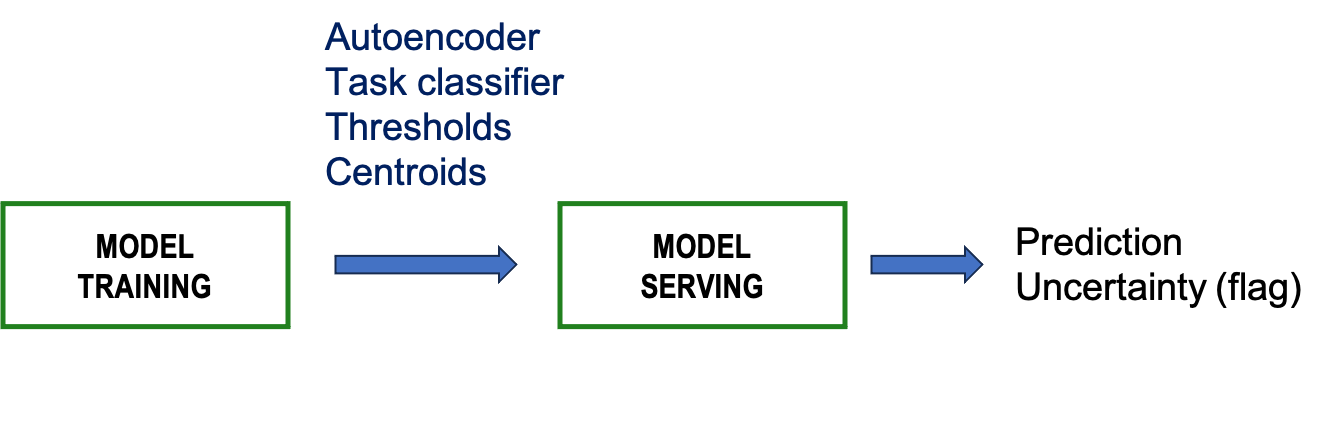}    \caption{Overall approach}
\label{fig:proposal} 
\end{figure} 

The overall approach is illustrated in Figure~\ref{fig:proposal}. Our approach follows a structured two-phase process. In the first phase, an autoencoder is trained to produce a well-organized latent space, alongside a classifier for the target task. At this stage, UQ methods are applied to determine optimal thresholds, which are set at a predefined quantile value—typically 0.99—ensuring that only the top 1\% of uncertainty scores are flagged as potential out-of-distribution data points. Additionally, latent space centroids, essential for distance-based assessments, are computed during this phase.

During model deployment, these key components, the classifier, the autoencoder, the thresholds and centroids, work together to deliver both predictions and uncertainty quantification, enhancing the reliability of the model in real-world conditions.


Furthermore, we adopt a flag-based approach to characterize uncertainty, as it is both practical and easily interpretable by domain operators. A green flag indicates that the input data falls within the training distribution, implying low prediction uncertainty, whereas a red flag signifies out-of-distribution data with high prediction uncertainty.



In the following sections, we present the implementation of our approach, including models and the UQ techniques.



\begin{figure*}[t]
  \centering
  \includegraphics[width=15cm]{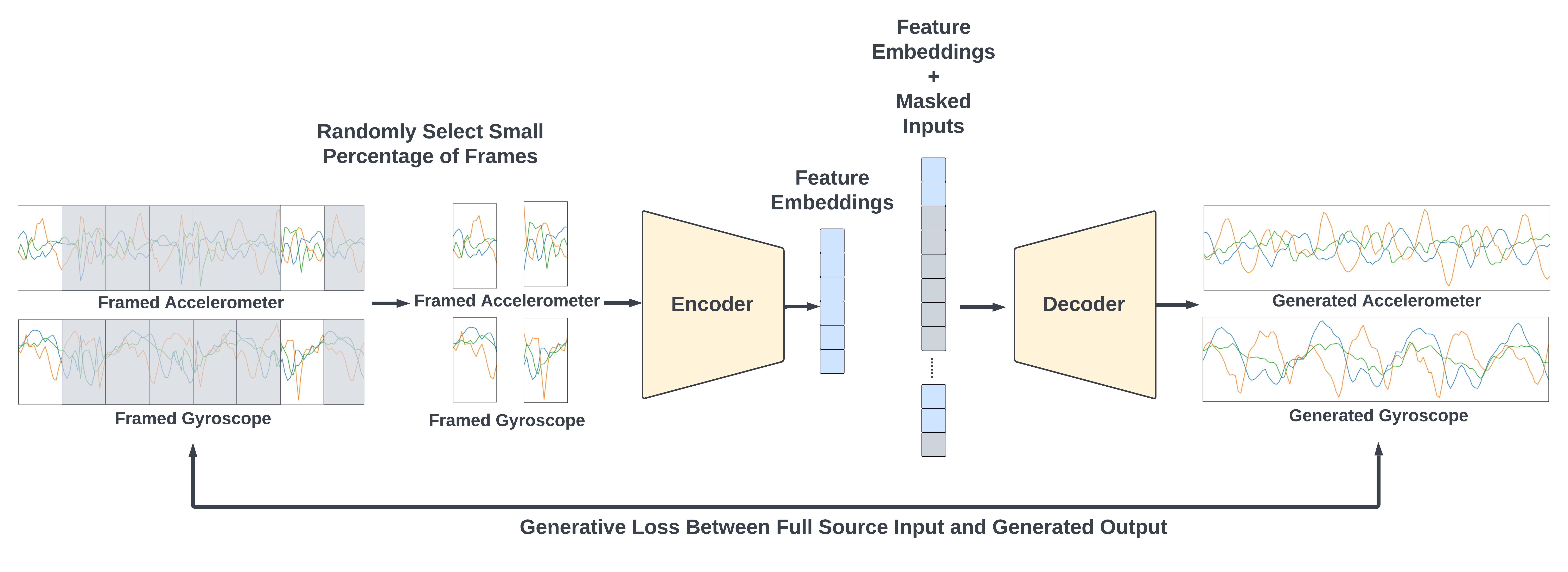}
  \caption{Model architecture combining MAE with a HAR Transformer \cite{Ek2024}. }
  \label{fig:tool}
\end{figure*}

\subsection{Autoencoder and domain models for HAR}

The first step in our approach involves constructing a robust latent space using an autoencoder. For the HAR domain, we achieve this by integrating a Masked Autoencoder (MAE) that employs a transformer at its core. Transformers excel at processing gyroscope and accelerometer signals, as their self-attention mechanisms effectively capture temporal dependencies and long-range relationships.

As illustrated in Figure~\ref{fig:tool}, MAE \cite{he2022masked} is a generative masking technique that learns the underlying structure of unlabeled data by reconstructing it from partial inputs. First, a linear projector transforms the input data into a fixed-size representation. Most of these representations are randomly masked using learnable mask tokens—vectors of the same length as the transformed input—which are optimized through back-propagation. Only a small subset of the unmasked frames is fed into the encoder. Finally, the encoder’s output is combined with the masked projections and passed to a lightweight decoder.

The loss function is typically defined as the Mean Squared Error (MSE) between the predicted and the original input values of the masked areas. It can be written as:

\begin{equation}
\mathcal{L}_{\text{MAE}} = \frac{1}{|M|} \sum_{i \in M} \| \mathbf{p}_i - \hat{\mathbf{p}}_i \|_2^2
\end{equation}

where $M$ is the set of masked inputs, $\mathbf{p}_i$ is the original frame value of the $i$-th masked frane, $\hat{\mathbf{p}}_i$ is the predicted frame value of the $i$-th masked frame, $\|\cdot\|_2$ denotes the L2 norm (Euclidean distance) and $|M|$ is the number of masked frames.

We adapted MAE for Human Activity Recognition (HAR) by building on its original implementation \cite{he2022masked}. We modified the input splitting (iSPL) module, responsible for dividing input data into patches, to process signals as a series of non-overlapping, independent frames, much like the way transformers operate on image patches. We also introduced sensor-specific masking tokens to better accommodate the unique characteristics of each sensor modality  \cite{Ek_2023}.

Our final model is a classifier designed to categorize human activities. The classifier, a fully connected neural network, processes latent representations extracted by the MAE to achieve accurate classification. A softmax layer normalizes the outputs into probabilities, ensuring a well-defined classification process. To mitigate overfitting, dropout is applied during training, improving model generalization. We actually explored three training approaches:
\begin{enumerate}
\item Freezing the MAE: The classifier is trained while keeping the MAE weights constant. This reduces computational cost and ensures stable feature extraction.
\item Fine-tuning the MAE: The MAE weights are updated during training, enhancing performance but increasing computational demand.
\item Random Initialization: The MAE is trained from scratch, requiring extensive learning and resources.
\end{enumerate}

Loss analysis revealed that fine-tuning and random initialization led to overfitting, while the frozen approach exhibited a smaller gap between training and validation losses, indicating superior generalization. By preserving pre-trained weights, this strategy curbs overfitting, streamlines training with fewer parameters to update, and leverages pre-learned features to boost classification accuracy without extensive retraining.



\subsection{Reconstruction loss}

\begin{figure}[h!]
\centering        
\includegraphics[width=\linewidth]{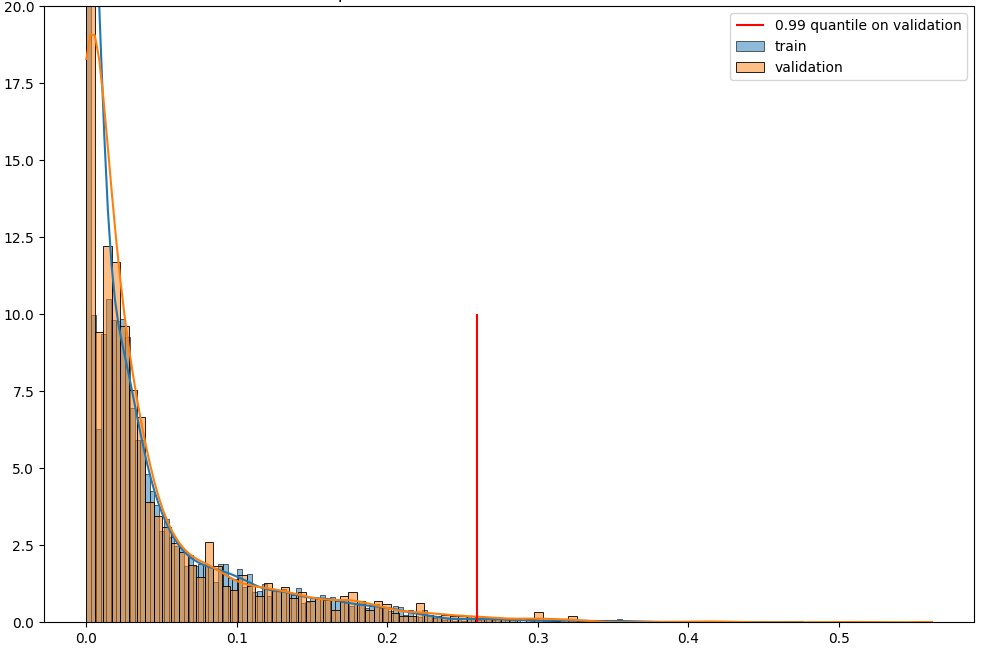}  
\caption{Reconstruction loss distribution for the training (blue) and validation (red) datasets, with the red line indicating the 0.99 quantile in the validation set.}
\label{fig:reco} 
\end{figure}

At the end of the MAE training phase, we evaluated the reconstruction loss on both training and validation datasets to establish a threshold that differentiates in-scope from out-of-scope data. Any data point with a loss exceeding this threshold is considered poorly represented in the training set and thus unreliable for prediction. As shown in Figure~\ref{fig:reco}, the loss values for both datasets are nearly identical and close to zero, indicating strong generalization and effective training. We defined the threshold using the 0.99 quantile, as indicated by the vertical line in the figure.

Note that one challenge in computing the reconstruction loss is that the MAE requires a minimum batch size of 8 consecutive frames, limiting the minimum signal duration. Additionally, small mini-batches can produce noisy loss values, as the model was trained with a batch size of 256. To mitigate this, we used sequential batches of 16×8 frames, balancing signal length and loss stability based on comparative loss distribution analysis.


\subsection{Distance calculation}

During the training phase, we also computed the distances between data points to establish a threshold that distinguishes in-scope from out-of-scope data. This evaluation was conducted on both the training and validation datasets. The low-dimensional representation produced by the MAE forms an effective vector space for distance calculations. This space preserves the data’s underlying structure, ensuring that similar inputs cluster closely while distinct ones remain well-separated. We actually applied clustering to the training data, using K-means—a widely used algorithm that partitions data into \( k \) clusters based on proximity to cluster centroids. The algorithm iteratively minimizes within-cluster variance by updating centroids as follows:


\begin{enumerate}
    \item Randomly initialize \( k \) centroids.
    \item Assign each data point to the nearest centroid based on Euclidean distance.
    \item Update centroids as the mean of assigned points.
    \item Repeat steps 2 and 3 until convergence or minimal centroid movement.
\end{enumerate}

For our dataset, the optimal number of clusters exceeded the product of body positions and activity classes, ensuring fine-grained segmentation that captures intra-class and positional variations, improving distance-based analysis.


We investigated alternative distance metrics—specifically, Mahalanobis distance, which accounts for feature correlations, and cosine similarity, which measures vector orientation. However, within the K-means framework, neither metric significantly outperformed the standard Euclidean distance. Ultimately, K-means using Euclidean distance, combined with an appropriately selected number of clusters, most effectively captured the data variations.

\begin{figure}[h!]
\centering        
\includegraphics[width=\linewidth]{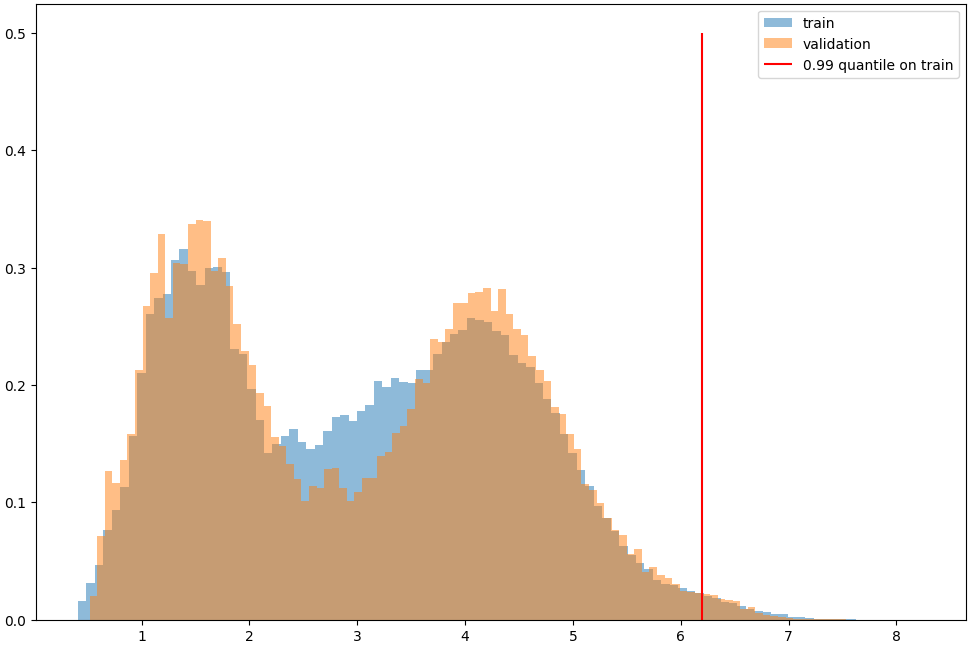}  
\caption{Distribution of distances to the nearest clusters for the training (blue) and validation (orange) datasets, with the red line indicating the 99\% quantile in the validation set. }
\label{fig:distance} 
\end{figure} 

Figure~\ref{fig:distance} illustrates the distribution of distances to the nearest clusters for both the training and validation datasets, along with the 99\% quantile from the validation set. This visualization highlights the spread of distances across datasets and the threshold used for outlier detection.

\subsection{Monte Carlo Dropout}

We then leveraged the validation set to compute the variance across multiple stochastic predictions generated via Monte Carlo Dropout. This metric was used to set a threshold for distinguishing between in-scope and out-of-scope data. A high variance indicates inconsistent predictions, suggesting increased uncertainty, while a low variance reflects stable and confident outputs.

\begin{figure}[h!]
\centering        
\includegraphics[width=\linewidth]{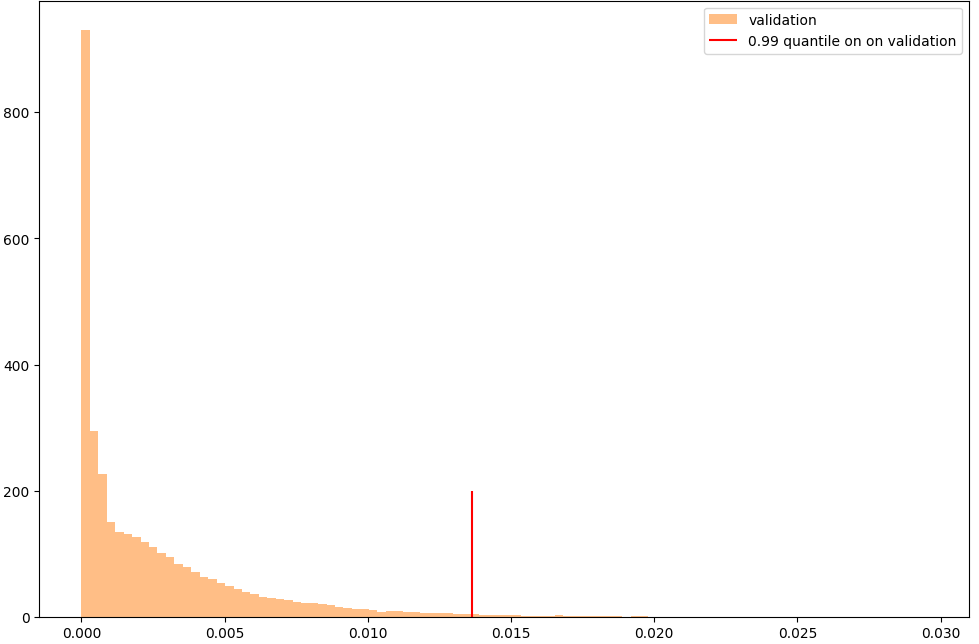}  
\caption{Variance distribution of dropout-induced predictions on the validation dataset, with the red line indicating the 99\% quantile.}
\label{fig:mcd} 
\end{figure}

In practice, the model was executed 100 times on the validation set with dropout enabled, yielding a comprehensive distribution of prediction variability. The threshold was defined as the 99\% quantile of this distribution, ensuring that nearly all reliable predictions fall below this level. Figure~\ref{fig:mcd} presents this distribution alongside the selected threshold, providing a clear visualization of the model's uncertainty. This approach not only enhances overall reliability but also systematically identifies ambiguous cases for further analysis

\begin{figure*}[t]
  \centering
  \includegraphics[width=16cm]{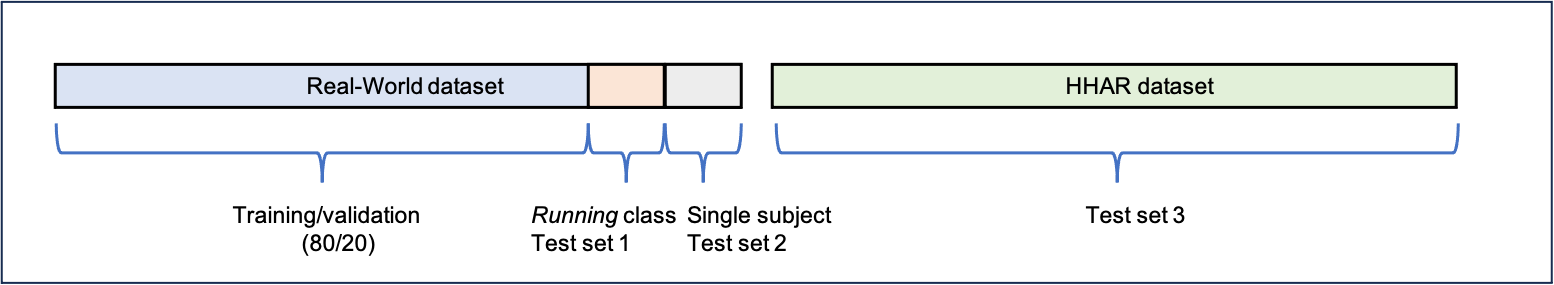}
  \caption{Training, validation and test datasets. }
  \label{fig:test}
\end{figure*}

\subsection{Cascading approach}

To effectively integrate UQ methods, we adopt a sequential strategy to reconcile differing outcomes. The methods are organized in order of increasing sensitivity and computational demand, guiding their application in the following order: 
\begin{enumerate}
    \item Reconstruction Loss: Applied first, this method labels a data point as out of scope if its reconstruction loss exceeds a predefined threshold. Otherwise, the point proceeds to the next step.
    \item Distance-based Methods: The point is evaluated against distance metrics (e.g., nearest cluster center). If the distance exceeds a threshold, it is labeled as out of scope; otherwise, it moves to the final step.
    \item Monte Carlo Dropout: Finally, the point undergoes MC Dropout evaluation. If its prediction variance is high, indicating uncertainty, it is labeled as out of scope.
\end{enumerate}

This cascading approach combines the strengths of each method. By applying techniques sequentially—using more expensive methods last—we effectively manage uncertainty and make informed decisions. This strategy improves outlier detection while reducing false positives and negatives, ensuring accurate classification through a comprehensive evaluation.

\section{Experiments}

\subsection{Selected datasets for training and testing}

We used two open HAR datasets: RealWorld \cite{sztyler2016onbody} and HHAR \cite{blunck2015hhar}. RealWorld comprises 125 hours of accelerometer and gyroscope data from 15 subjects across seven device configurations (Samsung Galaxy S4 and LG G Watch Z), covering eight activities. In contrast, HHAR provides 4.5 hours of recordings from 9 participants using multiple Android devices, covering six activities. Together, these datasets capture heterogeneous learning environments by incorporating diverse devices and user variations, enabling the construction of realistic evaluation scenarios.

As illustrated in Figure~\ref{fig:test}, these datasets were used to create a training set, a validation set, and three test sets. The training and validation sets were derived from RealWorld, with the "Running" class and one subject deliberately excluded. The training-validation split follows the common 80/20 ratio. This intentional omission allows the evaluation of different types of distribution shifts.

To assess label shift detection, we created a test set containing data from the excluded "Running" class, which was not seen during training by either the MAE or the classifier. This dataset, referred to as [Unseen Class], is sourced from RealWorld. The choice of this class was guided by its distinct separation from others in the embedding space.

Similarly, to evaluate covariate shift detection, we constructed a test set using data from the excluded subject, referred to as [Unseen Subject], also sourced from RealWorld. While it contains the same activity classes as the training and validation sets, its origin from a different individual introduces a shift in input feature distribution, despite unchanged labels.

To evaluate domain shift detection, we defined a test set based on HHAR, referred to as [Unseen Dataset]. This dataset significantly differs from RealWorld, introducing domain shift through variations in activity classes, sensor placement, and data collection procedures (as it would be defined in software engineering \cite{4724590}). While some activities overlap between the two datasets, others are entirely new, and even shared activities may differ due to contextual factors or participant instructions. Additionally, differences in sensor placement result in substantial discrepancies in recorded signals, even for similar activities.

Lastly, the validation set serves as a baseline for performance metrics. Since it shares the same distribution as the training set, it enables benchmarking under ideal conditions and facilitates direct comparisons with test scenarios affected by distribution shifts.

\subsection{Metrics}





To evaluate the model's ability to correctly assign an appropriate uncertainty flag (green or red), we introduce the \textit{Uncertainty Accuracy (UA)} metric. This metric quantifies discrepancies between the classifier’s predictions and its uncertainty estimates by identifying cases where:

\begin{itemize}
\item The uncertainty estimation method considers a data point \textit{in scope} (i.e., reliable), yet the classifier misclassifies it.
\item The uncertainty estimation method considers a data point \textit{out of scope} (i.e., anomalous), but the classifier correctly classifies it.
\end{itemize}

Uncertainty Accuracy is computed using the following formulas:
\begin{equation}
    \begin{aligned}
        \text{NumberOfInconsistent} &= \#(\text{OutOfScope and CorrectClass}) \\
        &\quad + \#(\text{InScope and IncorrectClass})
    \end{aligned}
\end{equation}

\begin{equation}
    \begin{aligned}
        \text{NumberOfConsistent} &= \#(\text{InScope and CorrectClass}) \\
        &\quad + \#(\text{OutOfScope and IncorrectClass})
    \end{aligned}
\end{equation}

Finally, the definition of Uncertainty Accuracy (UA) is the following:
\begin{equation}
    \text{UA} = 1 - \frac{\text{NumberOfInconsistent}}{\text{NumberOfInconsistent} + \text{NumberOfConsistent}}
\end{equation}

where \# denotes the count of instances meeting the specified conditions.

The UA metric highlights the crucial need for consistency between the model (classifier) and its uncertainty estimation methods. In an ideal system, these components should work in harmony to minimize discrepancies. A low UA indicates that the uncertainty quantification framework fails to accurately assess the performance of the model - essentially, it struggles to determine whether a prediction is reliable. Conversely, a high UA signifies strong alignment, where the uncertainty quantification framework effectively captures the model's behavior, leading to more robust and dependable predictions.

\subsection{Results}

As a reminder, the goal of these experiments is to evaluate our approach's ability to provide a reliable quantification of predictions. We do not focus on the accuracy of the predictions (which, in our case and generally in HAR, exceeds 95\% on the validation set). Therefore, in Table I, which presents the results, we emphasize the UA metric across various types of shifts.

The first key takeaway from these results is that, as expected, the different methods respond differently to each type of shift. Notably, each tested shift is best handled by a different approach. Specifically, the reconstruction-based method performs best for label shift (unseen class), the MC dropout method is most effective for covariate shift (unseen subject), and the distance-based method is best in handling domain shift (unseen dataset). In all these cases, the label correction rate (green or red) is between 70\% and 80\%.

First, the reconstruction-based method performs well on unseen data and unseen classes, as these new samples remain within a relatively similar distribution (corresponding to label shift and covariate shift), with comparable devices and ways of wearing them. However, it fails to handle domain shift, where the distribution differs significantly due to new devices and sensor placements. Unsurprisingly, it performs well on the validation set, as it was involved in the MAE training.

Next, it is clear that the distance-based method fails entirely to handle a new class, as classes are not fully separated in the latent space, even after the clustering performed by the MAE. However, it proves effective in handling a new user (covariate shift) and a new dataset (domain shift), where the new data is distinctly separated from the existing data in the latent space. As expected, this method works well with the validation set since its threshold was computed using this set (and the training set). 

Finally, the MC dropout method struggles with label shift (unseen class) and domain shift (unseen dataset). This can be attributed to the model's inability, as trained, to recognize new classes or novel conditions (such as a new user or different sensor placement), even with hyperparameter regularization. However, this regularization helps mitigate covariate shift more effectively (unseen subject). As for the other methods, this method works well with the validation set since its threshold was computed using this set (and the training set). 

The second particularly important lesson is that the cascading use of the three methods yields the best results in two of the three shift cases. This shows that uncertainty tends to decrease rather than increase when multiple methods are combined. This can be explained by the first observation: since the methods perform well in different scenarios, they can therefore be complementary.
In the case of label shifts and covariate shifts, the improvements over the best individual method are small or even slightly negative. However, it is important to note that one cannot know in advance what type of shift will be encountered. The use of cascading allows both shifts to be handled in an (almost) optimal way. Finally, it appears that the most significant improvement is achieved for the domain shift. Here, we can see that the methods complement each other strongly, resulting in an increase of almost 10\%. Here again, this can be explained by the fact that the methods focus on different types of shifts, all of which are present when using the HHAR dataset.

Table II presents the execution times of the different methods. We provide the times for 10K samples to smooth out variations as much as possible and offer a representative overview of these execution times. It appears that all three methods run in less than a hundred of a second on the NVIDIA L4 GPU. It is under a tenth of a second on high range smartphones. This makes this approach suitable for many pervasive applications, particularly in the industrial and wellness sectors.

\begin{table*}[t]
    \centering
    \setlength{\tabcolsep}{10pt}
    \begin{tabular}{lcccc}
        \hline
        \textbf{Method} 
            & \textbf{Unseen Class} 
            & \textbf{Unseen Subject} 
            & \textbf{Unseen Dataset} 
            & \textbf{Validation} \\
        \hline
        Reconstruction Loss & 77.00 & 71.18 & 54.73 & 79.64 \\
        Distance-based      & 70.19 & 72.99 & 70.29 & 80.25 \\ 
        Monte-Carlo Dropout & 20.92 & \textbf{73.87} & 39.92 & \textbf{80.57} \\
        Cascade Method      & \textbf{78.52} & 71.39 & \textbf{80.25} & 78.66 \\
        \hline
    \end{tabular}
    \caption{Performance of uncertainty estimation across different types of distribution shifts. 
    Bold values denote the best performance for each shift type.}
    \label{tab:performance_results}
\end{table*}

\begin{table}[h]
    \centering
    \begin{tabular}{|c|c|c|c|c|}
        \hline
        Method & Reconst. & Distance & MCD & Total \\
        \hline
        Exec. Time (s) & 226.0 & 2.8 & 380.0 & 608.8 \\
        \hline
    \end{tabular}
    \caption{Execution time (in seconds) for each method on 10K samples.}
    \label{tab:execution_time}
\end{table}


Let us now discuss the threshold settings in these experiments. Setting the 99\% quantile as a threshold is a conservative choice. In cases where distribution shifts are uncertain, such as in the [Unseen Subject] and validation datasets, Uncertainty Accuracy slightly increases. This suggests that while a strict threshold minimizes the impact of uncertain data, it may also reject correctly classifiable instances, leading to a marginal rise in Uncertainty Accuracy.

Selecting an alternative threshold without insights from test data is challenging. One approach is to adjust the out-of-scope threshold based on method sensitivity, while another aims to maximize validation accuracy by setting a middle-of-scope threshold. Our experiments indicate that shifting the out-of-scope threshold has little impact on datasets that are clearly out of distribution ([Unseen Class], [Unseen Dataset]), suggesting the methods are already effective at identifying such cases. However, in datasets where the distinction is less clear, such as [Unseen Subject], threshold adjustments behave similarly to those on the validation set. This highlights the importance of understanding the nature of the shift when fine-tuning these thresholds.

\section{Conclusion}

In this paper, we have introduced a framework that integrates UQ methods to effectively handle various types of distributional shifts by identifying most incorrect predictions. Experiments have demonstrated that cascading these methods generally outperforms using them individually. More importantly, this approach enables the management of different types of shifts with a single tool. By quantifying uncertainty, it helps users assess prediction reliability and make more informed risk management decisions.

These promising results pave the way for enhanced management of ML-based IoT services. Our primary objective, aligned with ongoing work, is to further strengthen the framework by integrating additional techniques, such as density-based analysis and Evidential Learning \cite{amini2020deep}. Preliminary results indicate that these additions could significantly improve uncertainty estimation across diverse types of distributional shifts.


\bibliographystyle{ieeetr} 
\bibliography{bibfile}

\end{document}